\documentclass{eptcs}
\usepackage[utf8]{inputenc}

\def\titlerunning{Hierarchical and Upstream-Downstream Composition of Stock and Flow Models}
\def\authorrunning{Meadows, Li and Osgood}

\usepackage{subcaption}
\usepackage[english]{babel}
\usepackage{amsthm}
\usepackage{amsmath, amssymb, mathrsfs,newtxmath}
\usepackage{hyperref}
\usepackage{cleveref}\usepackage{tikz,tikz-cd}
\usepackage{xypic}
\usepackage{graphicx}
\usepackage{amsfonts}

\usepackage{tikz}
\usetikzlibrary{shapes}
\usepackage{float}

\usepackage{xcolor}

\usepackage{aliascnt}
\newtheorem{theorem}{Theorem}[section]

\theoremstyle{definition}
\newtheorem{definition}[theorem]{Definition}

\theoremstyle{definition}
\newtheorem*{definition*}{Notational Conventions}

\theoremstyle{remark}
\newtheorem{remark}[theorem]{Remark}

\theoremstyle{definition}
\newtheorem{example}[theorem]{Example}

\theoremstyle{definition}
\newtheorem{construction}[theorem]{Construction}

\theoremstyle{definition}

\newcommand{\FinSet}{\operatorname{FinSet}}
\newcommand{\StockFlow}{\operatorname{StockFlow}}
\newcommand{\Fl}{\mathbf{Fl}}
\newcommand{\Ob}{\operatorname{Ob}}
\newcommand{\Mor}{\operatorname{Mor}}
\newcommand{\El}{\operatorname{El}}

\newcommand{\new}{\operatorname{new}}

\newcommand{\Flow}{\operatorname{Flow}}
\newcommand{\Inflow}{\operatorname{Inflow}}
\newcommand{\Outflow}{\operatorname{Outflow}}
\newcommand{\Stock}{\operatorname{Stock}}
\newcommand{\CTLink}{\operatorname{CTLink}}
\newcommand{\Link}{\operatorname{Link}}

\newcommand{\up}{\operatorname{up}}
\newcommand{\down}{\operatorname{down}}
\newcommand{\into}{\operatorname{in}}
\newcommand{\outto}{\operatorname{out}}
\newcommand{\ct}{\operatorname{ct}}
\newcommand{\fl}{\operatorname{fl}}
\newcommand{\st}{\operatorname{st}}

\title{Hierarchical and Upstream-Downstream Composition of \\ Stock and Flow Models}
\author{Nicholas Meadows, Xiaoyan Li, Nathaniel D Osgood }

\begin{document}

\maketitle

\begin{abstract}
The growing complexity of decision-making in public health and health care has motivated an increasing use of mathematical modeling.  An important line of health modeling is based on stock \& flow diagrams.  Such modeling elevates transparency across the interdisciplinary teams responsible for most impactful models, but existing tools suffer from a number of shortcomings when used at scale. Recent research has sought to address such limitations by establishing a categorical foundation for stock \& flow modeling, including the capacity to compose a pair of models through identification of common stocks and sum variables. This work supplements such efforts by contributing two new forms of composition for stock \& flow diagrams.  We first describe a hierarchical means of diagram composition, in which a single existing stock is replaced by a diagram featuring compatible flow structure. Our composition method offers extra flexibility by allowing a single flow in the stock being replaced to split into several flows totalling to the same overall flow rate.  Secondly, to address the common need of docking a stock \& flow diagram with another ``upstream'' diagram depicting antecedent factors, we contribute a composition approach that allows a flow out of an upstream stock in one diagram to be connected to a downstream stock in another diagram.  Both of these approaches are enabled by performing colimit decomposition of stock \& flow diagrams into single-stock corollas and unit flows.

\end{abstract}
\section{Introduction}
Mathematical modeling is increasingly being used to inform public health decision-making. Compartmental models constitute an important, well-established and widely-used form of such modeling.  Since the pioneering infectious diseases work of Ross \cite{ross1916application, ross1917application} and McKermack \cite{kermack1927contribution} a century, such models have offered insights on prevention, control and eradication of a broad set of health conditions, including a vast array of infectious diseases \cite{anderson1992infectious}. For the past four decades, stock \& flow models within the System Dynamics modeling have constituted an important and increasingly prominent component of this compartmental modeling tradition \cite{sterman2000business, forrester}. 

However, the long tenure of the use of such models -- and particularly their deployment during prolonged public health emergencies, such as that seen during the COVID-19 pandemic -- have highlighted the importance of supporting models evolving through contributions made across broad, interdisciplinary teams. As we have detailed elsewhere \cite{CompositionalStockFlowMfPHBook2022}, traditional approaches to compartmental modeling encounter numerous problems in supporting such teams, including those extending from their lack of support for modularity in modeling. Such lack of modularity raises great difficulties in ensuring a division of labour within large modeling teams, inhibits model evolution by undercutting the ability to reuse model subpieces developed by others, and reduces efficiency due to difficulty of sustainably sharing, reusing and evolving robust model subpieces. 

Past contributions from the authors and collaborators have recently advanced a diagram-centric approach to compartmental modeling founded in category theory \cite{AlgebraicStockFlow, CompositionalStockFlowMfPHBook2022}. This effort has contributed underlying theory for the categorical treatment of stock and flow (henceforth, stock \& flow) diagrams based on undirected wiring diagrams, structured cospans, presheaves employing schema categories representing the grammar for stock \& flow diagrams \cite{ACTStockFlow2022, CompositionalStockFlowMfPHBook2022}, drawing heavily on the pioneering work of Baez and students Pollard 
\cite{baezpollard2017}, Courser \cite{baezcourser2020}, Vasilakopoulou together with  Courser \cite{baez-courser-vasilakopoulou2022}.  But it has further contributed an implementation of the theory in the form of the StockFlow \cite{AlgebraicStockFlow} component of AlgebraicJulia \cite{AlgebraicJulia}, and the ModelCollab real-time collaborative graphical editor for assembling and composing such models \cite{CompositionalStockFlowMfPHBook2022}. This categorical approach to stock \& flow diagrams has demonstrated many benefits, including the capacity to compose models by identifying stocks and (separately) sum auxiliary variables \cite{AlgebraicStockFlow, CompositionalStockFlowMfPHBook2022}, a capacity to separate syntax characterizing diagrams from the semantics by which to interpret them \cite{CompositionalStockFlowMfPHBook2022}, and modular stratification of such diagrams \cite{CompositionalStockFlowMfPHBook2022}.

However, as identified by the authors in that work, there is a great need for expansions of the categorical foundation of the work, including by supporting additional types of composition of stock \& flow diagrams \cite{CompositionalStockFlowMfPHBook2022}.  This report responds to that call by providing a solution to two common needs in the use of stock \& flow models:  Firstly, we address the need to expand the scope of a diagram by expanding one area of the model in greater detail -- to ``zoom in", as it were, so as to provide greater specifics about the dynamics of a model region previously treated in an aggregate fashion.  Secondly, we formally characterize a means of composing an upstream stock \& flow diagram with a downstream such diagram by having the former flow into the latter.  While the model composition strategy characterized and implemented in our previous work \cite{AlgebraicStockFlow, CompositionalStockFlowMfPHBook2022} addressed the need to extend a diagram \textit{laterally} (gluing one stock \& flow diagram onto stocks and sum auxiliary variables of another), the approach contributed here extends a stock \& flow diagram by having it flow into a downstream diagram via an existing flow, and (separately) \textit{hierarchically}, replacing a given stock embedded within a surrounding model with the whole of a another ``compatible'' stock \& flow diagram.  These contributions draw on ideas from the whole-grain petri net approach of Kock \cite{kock2022whole} and achieving colimit decomposition of stock \& flow diagrams into corollas and unit flows.

The structure of this paper is as follows: In the next section, we describe the notation and terminology employed. Section \ref{section: elementary stock flow} introduces the simplified form of stock \& flow diagrams treated here, pushouts of such diagrams, and the key notions of a corolla and of a unit flow.  Section \ref{section: hierarchical stock flow} formalizes the concept of a hierarchical stock \& flow diagram, along the way introducing other key supporting concepts such as half flows and providing examples from stock \& flow diagrams inspired by practice. Section \ref{section: subdividing flows} further tackles a subtlety associated with practical use of this approach -- the need during  hierarchical substitution to subdivide what was a single flow out of the aggregate stock being replaced into multiple flows within the more detailed diagrams.  Section  \ref{section: upstream downstream composition} continues on to employ the constructions introduced earlier -- particularly corollas and unit flows -- to describe a technique for composing a pair of stock \& flow diagrams in an upstream-downstream fashion.  Within the context of such composition, an outflow from the upstream diagram flows into an inflow from the downstream diagram.  Within Section \ref{section: conclusion}, we place this work in context and offer some closing remarks.

\section{Elementary Stock \& Flow Diagrams}
\label{section: elementary stock flow}

\begin{definition}\label{def1.1}

We define by $\Fl$ the category freely generated by the graph:

\begin{equation}\label{eq1}
\xymatrix
{
& \Outflow \ar[dr]^{\outto} \ar[dl]_{up} & \\
\mathrm{Stock} &  & \Flow \\
&  \Inflow \ar[ur]_{\into} \ar[ul]^{\mathrm{down}} & \\
 & & \\
\mathrm{CTLink} \ar[uuu]^{\mathrm{st}} \ar[rr]_{\mathrm{ct}} &&  \mathrm{Link} \ar[uuu]_{\mathrm{fl}}  
}
\end{equation}

A \emph{primitive stock \& flow diagram} is a presheaf $X : \Fl  \rightarrow FinSet$ such that
$$X(\mathrm{in}), X(\mathrm{out}), X(\mathrm{ct})$$ are monomorphisms. This just says that the flows which appear as out-/in-flows are a subset of the collection of all flows, and the links which are connected to a stock are a subset of all links. 

A \emph{stock \& flow diagram} is a primitive stock \& flow diagram $X$, along with a continuous flow function $\phi_{f}: \mathbb{R}^{X(\mathrm{fl})^{-1}(f)} \rightarrow \mathbb{R}$ for each $f \in X(\Flow)$.

We write $(X, \phi_{\bullet})$ for the stock flow diagram with flow functions $\phi_{f} : f \in X(\Flow)$.  We introduce the necessary terminology below. 

Suppose that $(X, \phi_{\bullet}), (Y, \psi_{\bullet})$ are stock \& flow diagrams, and $F : X \rightarrow Y$ is a presheaf morphism. Suppose further that $F(\Flow)(f) = g$ (where $f \in X(\Flow)$, and $g \in Y(\Flow)$); then we get an induced map:
$$
F(\Link) : X(\fl)^{-1}(f) \rightarrow Y(\fl)^{-1}(g)
$$ 
and thus a linear map 
$$
F(\Link)^{*} : \mathbb{R}^{Y(\fl)^{-1}(g)} \rightarrow \mathbb{R}^{X(\fl)^{-1}(f) }
$$
given by precomposition: 
$$
F(\Link)^{*}(x) = x \circ F(\Link) 
$$

 We say that a presheaf morphism $F : X \rightarrow Y$ is a \emph{morphism of stock \& flow diagrams} $(X, \phi_{\bullet}) \rightarrow (Y, \psi_{\bullet})$ if for each nonempty $F(\Flow)^{-1}(g)$  
 \begin{equation}\label{eq0}
 \psi_{g} = \sum_{f \in F(\Flow)^{-1}(g)} \phi_{f}  \circ F(\Link)^{*} 
  \end{equation}

  
  Composition of morphisms of stock \& flow diagrams is composition of their underlying natural transformations. It can be shown that composing two natural transformations that obey Equation~\ref{eq0} yields a new natural transformation obeying Equation~\ref{eq0}. We thus obtain a category called $\StockFlow$.

 \end{definition}

\begin{remark}\label{rmk1.2}
The definition of primitive stock \& flow diagrams above is different from that of \cite{StockFlow}. Namely, we have modified the schema to allow for unit flows --- free hanging flows --- and links absent a source stock. These are essential for the colimit decomposition that plays a central role in defining hierarchical stock \& flow diagrams.

 The definition of morphisms of stock \& flow diagrams from \cite{StockFlow} is different as well. The authors of \cite{StockFlow} specify that Equation~\ref{eq0} holds for every  $g \in Y(\Flow)$, which implies that each map of stock \& flow diagrams must be surjective on stocks. The greater flexibility in our definition of morphisms of stock \& flow diagrams allows for the colimit decomposition of a stock \& flow diagrams central to this work.  
\end{remark}
\begin{example}\label{exam1.3}
The standard way in which we depict stock \& flow diagrams is as follows. 
\begin{enumerate}
\item{We depict stocks as boxes $\boxed{Stock}$ }
\item{We depict links as ordinary arrows $\longrightarrow$, pointing from a stock to a flow.}
\item{We depict flows as arrows $\Longrightarrow$ indicating the direction of the flow}
\item{Sources and sinks are depicted as clouds \scalebox{0.2}{
\begin{tikzpicture}
\node [cloud, draw,cloud puffs=10,cloud puff arc=120, aspect=2, inner ysep=1em] {};
\end{tikzpicture}
}}
\end{enumerate}
\begin{figure}[H]
    \centering
     \includegraphics[scale=0.3]{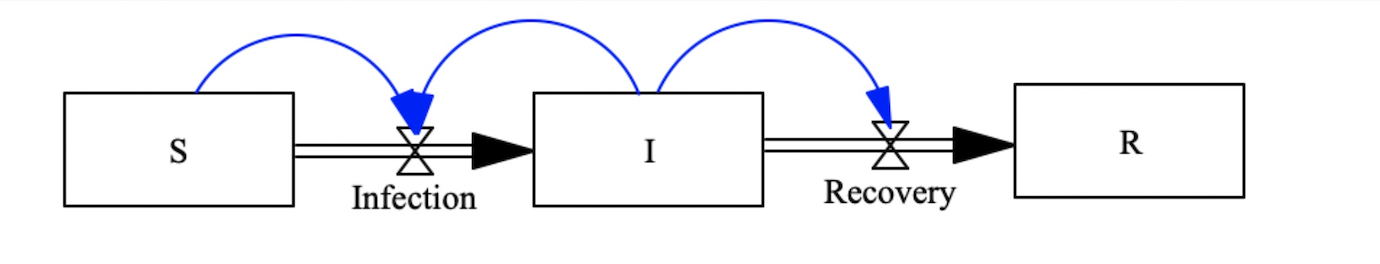}
    \caption{A simple stock \& flow diagram}
    \label{StockFlow_figure}
\end{figure}

\noindent
Figure \ref{StockFlow_figure} depicts a simple stock \& flow diagram with three stocks, two non-half-flows (connect to both an upstream stock and a downstream stock) and three links. 

\end{example}

\begin{theorem}\label{thm1.4}
Suppose that we have a diagram of stock \& flow diagrams 
\begin{equation}\label{exx}
\xymatrix{
(X, \phi_{\bullet}) \ar[d]_{s_{2}} \ar[r]_{s_{1}} & (Z, \kappa_{\bullet}) \\
(Y, \psi_{\bullet}) &  
}
\end{equation}
in which $s_{1}(\Flow) : X(\Flow) \rightarrow Z(\Flow), s_{2}(\Flow) : X(\Flow) \rightarrow Y(\Flow)$ are surjective. Consider the presheaf pushout 

$$
\xymatrix
{
X \ar[r]_{s_{1}} \ar[d]_{s_{2}} & Z \ar[d]^{p_{1}}\\
Y \ar[r]_{p_{2}} & Z \coprod_{X} Y
}
$$
and for each $f \in (Z \coprod_{X} Y)(\Flow)$, let 
$$ 
\gamma_{f} = \sum_{x \in (p_{2} \circ s_{2})(Flow)^{-1}(f)} \phi_{x} = \sum_{x \in (p_{1} \circ s_{1})(Flow)^{-1}(f)} \phi_{x}
$$

Then the pushout of Diagram~\ref{exx} is given by $(Z \coprod_{X} Y, \gamma_{\bullet})$.

\end{theorem}

\begin{definition}\label{def2.1}
A \emph{half flow} in a stock \& flow diagram is a flow that appears as an inflow or an outflow, but not both. A \emph{closed} stock \& flow diagram is one in which there are no half flows. 
\end{definition}

\begin{definition}\label{def1.5}
A unit flow is a stock \& flow diagram  $(X, \phi_{\bullet})$ such that it behaves on objects by:

\begin{equation}\label{eq1_4}
X(p) = 
\left\{
	\begin{array}{ll}
		* & \mbox{if } p = \Flow \\
        L & \mbox{if } p = \Link \\
        \emptyset & \mbox{if } p \neq  \Flow, \Link
	\end{array}
\right.
\end{equation}

where $*$ is the one point set. Essentially, it is a free-hanging flow with a number of free-hanging half links pointing to it.

 A \emph{corolla} is simply a stock \& flow diagram $(X, \phi_{\bullet})$ with $X(\Stock) = *$, the one point set. $X$ can be depicted as  
$$
\xymatrix
{
\, \, \, \, \, \, \  \, \, \, \, \, \, & \ar[dr]^{\outto} F_{\outto} \ar[dl]_{\up} &   \\ 
\, * &   & F_{\outto} \coprod F_{\into}\\
&  F_{\into} \ar[ur]^{\mathrm{in}} \ar[ul]^{\down} & \\
 & & \\
 L_{Ct} \ar[rr]_{\ct} \ar[uuu]^{\st} &  & \ar[uuu]_{\fl} L
}
$$

In this situation, the links are the disjoint union of (incoming) free-hanging links and links connected to (i.e., from) the stock, and the flows are a disjoint union of half inflows/outflows from the stock. 

\end{definition}

\begin{definition}\label{def1.6}

Let $S$ be a set. We write $\mathrm{Disc}_{S} = (\mathrm{Disc}_{S}, \emptyset)$ for the stock \& flow diagram with 

\begin{equation*}
\mathrm{Disc}_{S}(p) = 
\left\{
	\begin{array}{ll}
		S & \mbox{if } p = \Stock \\
		\emptyset & \mbox{if } \text{otherwise}
	\end{array}
\right.
\end{equation*}
Since there are no flows, there are no flow functions

\end{definition}

An example of such a discrete diagram is shown in Figure \ref{DISCRETE}.


Given a stock \& flow diagram $(X, \phi_{\bullet})$, there is a unique map $\mathrm{Disc}_{X(\Stock)} \rightarrow (X, \phi_{\bullet})$, which is the identity on stocks. This is indeed a morphism, since by definition Equation~\ref{eq0} only needs to hold for flows $f$ such that $F(\Flow)^{-1}(f)$ is nonempty.

\begin{definition}\label{def1.7}
Suppose that $(X, \phi_{\bullet})$ is a stock \& flow diagram. Given $s \in X(\Stock)$, the corolla associated with $s$, $\mathcal{C}_{s}$, is the stock \& flow diagram $(C_{s}, \psi_{\bullet})$ with $C_{s}$ depicted as:

$$
\xymatrix
{
&  \ar[dr]^{X(\outto)} X(\up)^{-1}( \{ s \}) \ar[dl]_{X(\up)} & \\
 \, \, \, \, \, \, \ * \, \, \, \, \, \,  &  &(X(\down)^{-1}(\{ s\})) \coprod  
 (X(\up)^{-1}(\{ s\}))
 \\ 
&  X(\down)^{-1}(\{ s\}) \ar[ur]^{X(\into)} \ar[ul]^{X(\down)} & \\
 & & \\
\ar[rr]_{X(\ct)} \ar[uuu]^{X(\st)} X(\st)^{-1}(\{ s\})  &  & \ar[uuu]_{X(\fl)} X(\fl)^{-1}X(\down)^{-1}(\{ s\}) \coprod X(\fl)^{-1}X(\up)^{-1}(\{ s\})
}
$$
and $\psi_{f} = \phi_{f}$ for $f \in  X(\down)^{-1}(\{ s\}) \coprod X(\up)^{-1}(\{ s\})$ and each of the morphisms $m \in \Mor(\Fl)$ $\mathcal{C}_{s}(m)$ being a restriction of $X(m)$.

The \emph{unit flow} $(U_{f}, \psi_{\bullet})$ associated with a given flow $f$ is precisely as in Equation \ref{eq0}, with $L$ replaced with $X(\fl)^{-1}(f)$. The unique flow function $\psi_{f}$ is just $\phi_{f}$.

\end{definition}

Essentially, the corolla associated with a stock $s$ consists of all inflows and outflows of that stock (as half flows or self-flows), as well as all links pointing to them.  It is important to note that those links originating from a stock distinct from $s$ appear as `half links', which don't have a source stock.

\begin{example}\label{exam1.8}
A unit flow consists of free-hanging links pointing to single flow, as in Figure \ref{unitflow}. 

\begin{figure}[tp]
\centering
\begin{subfigure}[t]{0.4\linewidth}
    \centering
    \includegraphics[width=0.6\linewidth]{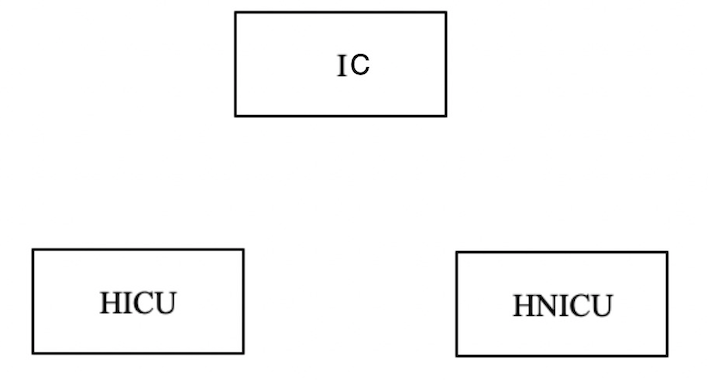}
    \caption{The discrete diagram $\mathrm{Disc}_{\{ \text{IC}, \text{HICU}, \text{HNICU} \}}$}
    \label{DISCRETE}
\end{subfigure}\hfil
\begin{subfigure}[t]{0.2\linewidth}
    \centering
    \includegraphics[width=\linewidth]{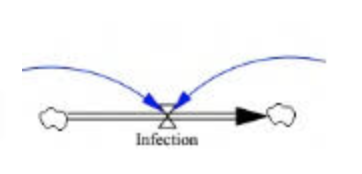}
    \caption{A unit flow}
    \label{unitflow}
\end{subfigure}\hfil
\begin{subfigure}[t]{0.4\linewidth}
    \centering
    \includegraphics[width=\linewidth]{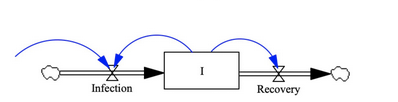}
    \caption[b]{A corolla}
    \label{corolla}
 \end{subfigure}
\caption{Examples of a discrete diagram, a unit flow and 
a corolla}
\end{figure}




On the other hand, the corolla associated with the stock `I' from the diagram in Example \ref{exam1.3} is shown in Figure \ref{corolla}.




\end{example}

\begin{construction}\label{con1.8}
Suppose that $(X, \phi_{f})$ is a stock \& flow diagram.
Let $\mathcal{U}_{f} = (U_{f}, \psi_{\bullet})$ be the unit flow associated with $f$ in $\StockFlow$. There is an 
obvious inclusion morphism, which we denote $\mathfrak{F}_{f}^{X} :  \mathcal{U}_{f} \rightarrow X$. 

We will define a subcategory $\El_{/X}$ of $\StockFlow$ as follows. We have 
\begin{align*}
\Mor(\El_{/X}) = \{ \mathfrak{F}_{f}^{\mathcal{C}_{s}} : \mathcal{U}_{f} \rightarrow \mathcal{C}_{s}   : s \in X(\Stock), f \in \mathcal{C}_{s}(\Flow) \} \\
\Ob(\El_{/X}) = \{ \mathcal{C}_{s} : s \in X(\Stock) \} \cup \{ \mathcal{U}_{f} : f \in X(\Flow)  \}
\end{align*}

\end{construction}

\begin{theorem}\label{corolladecomp_theorem}
X is the colimit of the inclusion functor $\El_{/X} \rightarrow \StockFlow$. 
\end{theorem}

\begin{example}\label{exam1.10}
The colimit decomposition of the stock \& flow diagram from Example \ref{exam1.3} is:

\begin{figure}[H]
    \centering
 \includegraphics[width=0.9\textwidth]{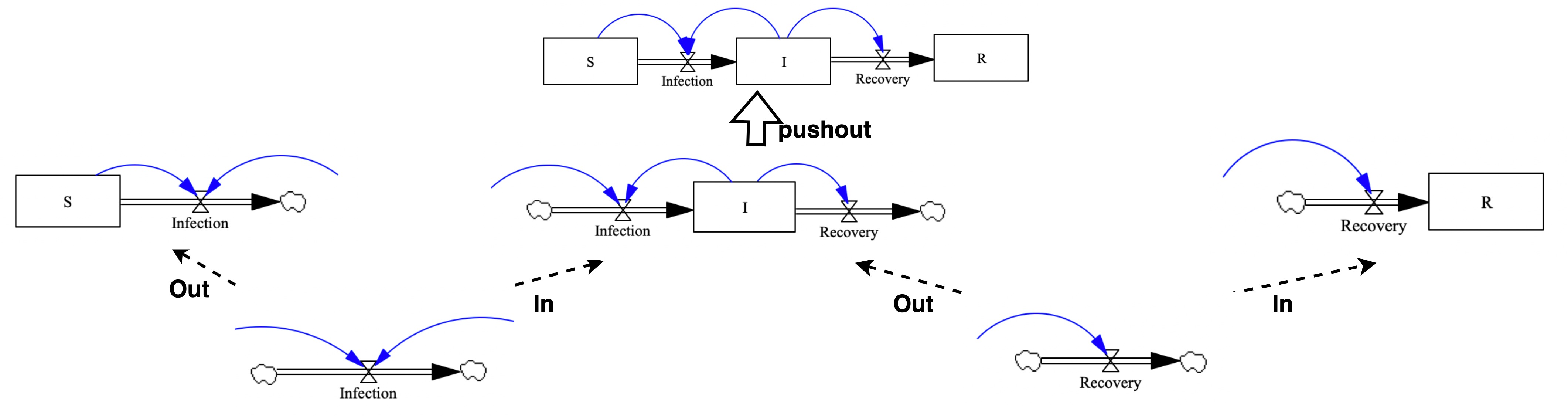}
    \caption{An example of the colimit decomposition into corollas}
    \label{corolladecomp_figure}
\end{figure}

The middle row depicts the corollas associated with each stock. 

\end{example}

\section{Hierarchical Stock \& Flow}
\label{section: hierarchical stock flow}

In this section, we want to formalise the idea of a \emph{hierarchical stock \& flow diagram}. Roughly speaking, this formalises the idea of replacing each stock in a given stock \& flow diagram with an entire stock \& flow diagram, in a way that respects the out/inflows and link structure associated with a given stock. More precisely, we want to replace each corolla with an entire stock \& flow diagram, in a way that respects \emph{half flows} and their associated link structure. This is achieved by computing a colimit of a diagram of a specific shape, indexed by $\El_{/X}$.

\begin{example}\label{exam2.2}

Suppose we want to replace a corolla $\mathcal{C}_{s} = (C_{s}, \psi_{\bullet})$ of a stock \& flow diagram $(X, \phi_{\bullet})$ with a more complicated stock \& flow diagram in a manner that preserves the flows in and out of the corolla. Informally, this operation involves choosing a closed stock \& flow diagram  $(Y, \gamma_{\bullet})$, which will be substituted for the stock of $\mathcal{C}_{s}$, and then attaching the half flows in and out of $\mathcal{C}_{s}$ to stocks of $(Y, \gamma_{\bullet})$ instead.

As an example, we may want to replace the corolla of $I$ in the diagram in Figure \ref{StockFlow_figure} with the diagram

\begin{figure}[H]
    \centering
 \includegraphics[width=0.3\textwidth]{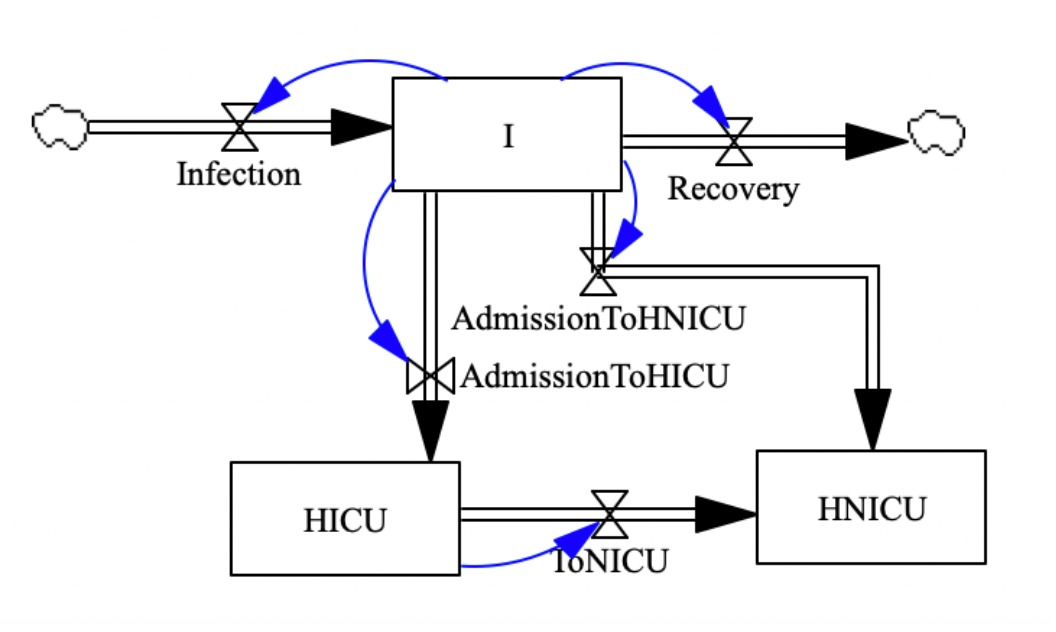}
    \caption{A closed stock \& flow diagram}
    \label{closed}
\end{figure}

\end{example}

\begin{construction}\label{con2.3}

We want to explain in mathematical terms the construction of Example \ref{exam2.2}. Fix a corolla $\mathcal{C}_{s}$, along with a closed stock \& flow diagram $(Y, \gamma_{\bullet})$ which we want to substitute for the stock $s$ of $\mathcal{C}_{s}$.

We will write $\Fl'$ for the subcategory of $\Fl$ generated by:

$$
\xymatrix
{
& \Outflow \ar[dr]^{\outto}  & \\
 &  & \Flow \\
&  \Inflow \ar[ur]_{\mathrm{in}}  & \\
 & & \\
\CTLink \ar[rr]_{\ct} &  &  \Link \ar[uuu]_{\fl}
}
$$

 We write $C_{s}|_{\Fl'}$ for restriction of the underlying presheaf of $C_{s}$ to $\Fl'$ and $[0]$ for the trivial category with one object. Let us solve the extension problem: 
\begin{equation}\label{eq3}
\xymatrix
{
\Fl' \coprod [0] \ar[rrr]^{(C_{s}|_{\Fl'}, Y(\Stock))} \ar[d]_{(\mathrm{inc}, \Stock)} &&&  \FinSet \\
\Fl \ar@{.>}[urrr]_{D} &
}
\end{equation}
where $\mathrm{inc}$ is inclusion functor.

Together with the flow functions from $\mathcal{C}_{s}$, we get a new stock \& flow diagram $(D, \psi_{\bullet})$. The solution of this extension problem is, informally, a choice of way of connecting half flows of the corolla to the stocks of  $(Y, \gamma_{\bullet})$. 

The diagram replacing the corolla can be computed as the pushout: 
 
\begin{equation}\label{eq4}
\xymatrix
{
\mathrm{Disc}_{Y(\Stock)} \ar[d] \ar[r] &(D, \psi_{\bullet})  \ar[d]^{p} \\
(Y, \gamma_{\bullet}) \ar[r] & (E, \omega_{i})
}
\end{equation}

Note that since $D(\Flow) = \mathcal{C}_{s}(\Flow)$, we have a map $\mathcal{U}_{f} \rightarrow (D, \psi_{*}) $ which sends the unique point of $\mathcal{U}_{f}$ to $f$.
Postcomposing with $p$, we get a map 
\begin{equation}\label{eq5}
\mathfrak{G}_{f}^{\mathcal{C}_{s}} : \mathcal{U}_{f} \rightarrow (E, \omega_{\bullet})
\end{equation}

\end{construction}

\begin{example}\label{exam2.4}
In the case of Example \ref{exam2.2}, Construction \ref{con2.3} can be depicted as the pushout:

\begin{figure}[H]
    \centering
 \includegraphics[scale=.1]{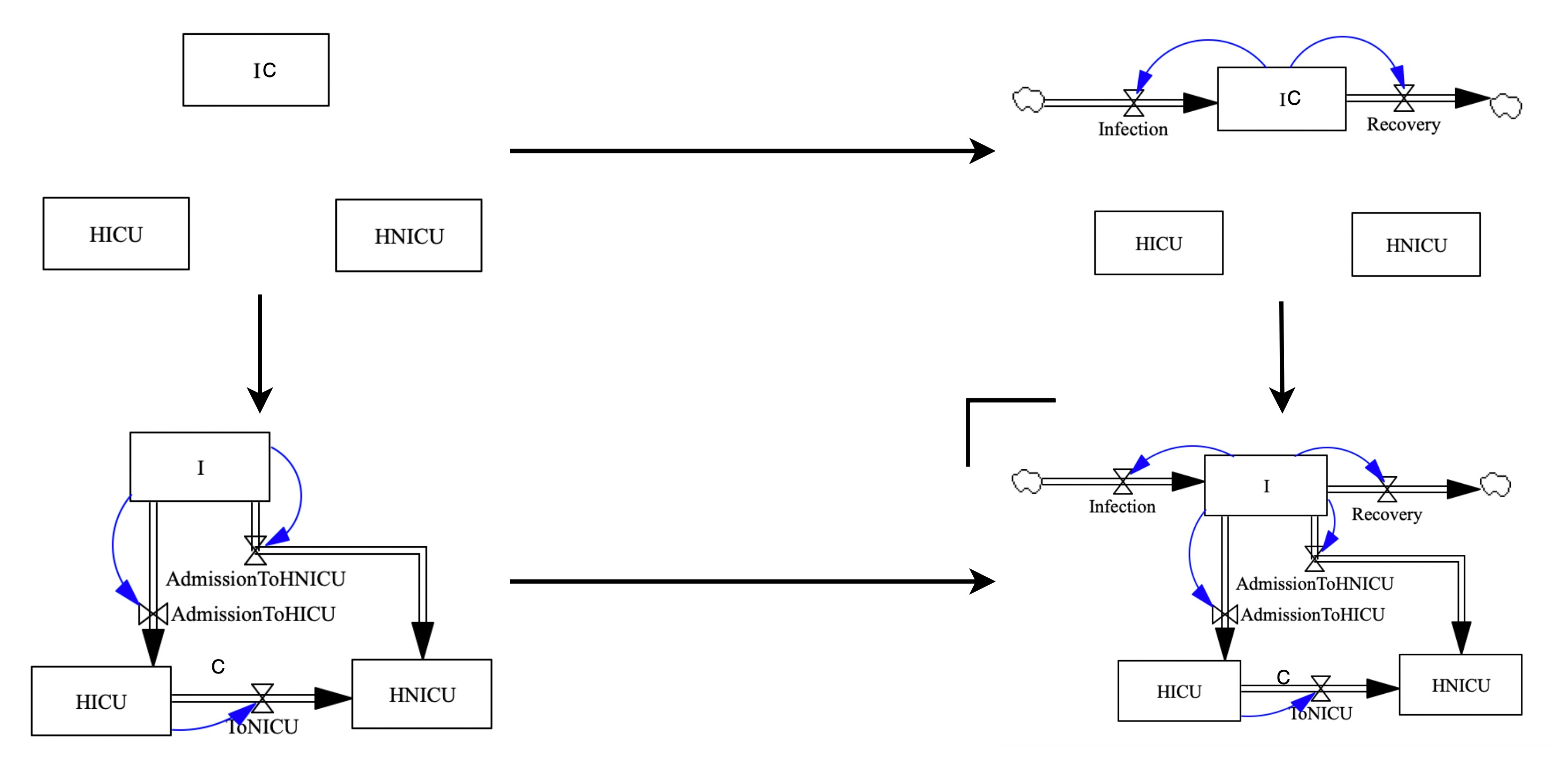}
    \caption{An example of Construction \ref{con2.3}}
    \label{hierarchical}
\end{figure}

\end{example}

\begin{definition}\label{def2.5}
A \emph{hierarchical stock \& flow diagram} $F : \El_{/X} \rightarrow \StockFlow$ is a diagram such that
\begin{enumerate}
\item{It takes corollas to pushouts of diagrams of the form diagram \ref{eq4} }
\item{$\mathcal{U}_{f} \mapsto \mathcal{U}_{f}$}
\item{$\mathfrak{F}_{f}^{\mathcal{C}_{s}} \mapsto \mathfrak{G}_{f}^{\mathcal{C}_{s}}$  (from map \ref{eq5})}
\end{enumerate}

\end{definition}

\begin{example}\label{exam2.7}
We can replace hierarchical stock \& flow diagrams $F: El_{/X} \rightarrow \StockFlow$ with more complicated stock \& flow diagrams by computing the colimit of $F$. As an example, after subdividing the flow 'Recovery' into two different flows $Recovery', RecoveryFromHNICU$ (something we describe more formally in the next section) we can replace the corolla $I$ in the diagram of Example \ref{exam2.2} by computing the colimit:

\begin{figure}[H]
    \centering
 \includegraphics[width=0.9\textwidth]{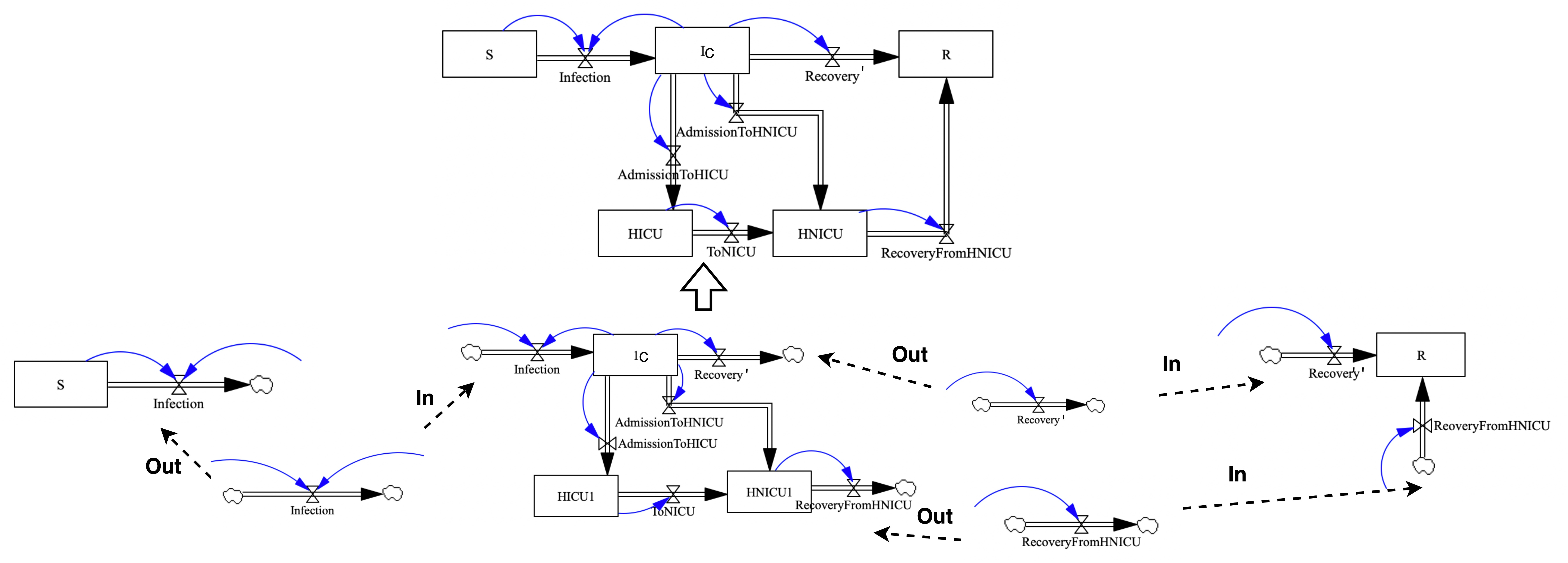}
    \caption{A hierarchical stock \& flow diagram}
    \label{corolladecomp}
\end{figure}

\end{example}

\section{Subdividing Flows}
\label{section: subdividing flows}

In this section, we explain how to subdivide an inflow or outflow of a corolla. Intuitively, this is an important part of constructing hierarchical stock \& flow diagrams; the inflow (outflow) of a stock will be divided into various flows coming from (going to) multiple stocks.

\begin{example}\label{exam3.1}
In this example, we will show how to further decompose a corolla in terms of stock \& flow diagrams with a single flow.
Let $(C_{s}, \psi_{\bullet})$ be a corolla. Without loss of generality, let us assume that we have chosen an inflow
$f$ of $C_{s}(\Flow)$. Consider the primitive stock \& flow diagram $C_{s}^{f}$,
depicted according to the schema of Definition \ref{def1.1} as:

\begin{equation}\label{eq2b}
\xymatrix
{
& \emptyset \ar[dr]^{\outto} \ar[dl]_{\up} & \\
\, * &  & \ * \\
& \, * \ar[ur]_{\into} \ar[ul]^{\mathrm{down}} & \\
 & & \\
C_{s}(\ct)^{-1}C_{s}(\fl)^{-1}(f)  \ar[uuu]^{\st} \ar[rr]_{\ct} &    & C_{s}(\fl)^{-1}(f) \ar[uuu]_{C_{s}(\fl)}
}
\end{equation}

Let $[1]$ be the walking arrow category. Each corolla can be expressed as the colimit of the diagram 
\begin{equation}\label{corolladecomposition}
\Phi:  [1] \coprod_{[0]} [1] \cdots \coprod_{[0]} [1] \rightarrow \StockFlow,
\end{equation}  
Where the domain represents a category which is an iterated pushout with a summand for each element of $C_{s}(\Flow)$.  $\Phi$ sends the summand corresponding to flow $f \in C_{s}(\Flow)$ to the unique morphism $\mathrm{Disc_{\{ *\}}} \rightarrow \mathcal{C}_{s}^{f}$.

\end{example}

With the above results for primitive stock \& flow diagrams in hand, we now address the case of stock \& flow diagrams, which are additionally equipped with flow functions.

\begin{construction}\label{con3.2}

Let $n$ be a fixed positive integer. We are going to explain how to split a chosen flow $f$ into $n$ different flows in a stock \& flow diagram.
Consider the functor of primitive stock \& flow diagrams induced by the identity on each summand from a iterated pushout of $n$ copies of $C^{f}_{s}$:
\begin{equation}
\label{diagonalMap}
(\coprod_{i=1}^{n} C^{f}_{s}) \rightarrow C^{f}_{s}
 \end{equation}

We wish to extend this map of primitive stock \& flow diagrams \ref{diagonalMap} to a map of stock \& flow diagrams $\mathcal{C'}_{s}^{f} := (\coprod_{i=1}^{n} C^{f}_{s} , \gamma_{\bullet}) \rightarrow (C_{s}^{f}, \psi_{\bullet})$.  Consider the linear map induced by diagram \ref{diagonalMap}:
$$
\Phi(\Link)^{*} : (\mathbb{R}^{C_{s}(\mathrm{fl})^{-1}(f)}) \rightarrow (\mathbb{R}^{C_{s}(\mathrm{fl})^{-1}(f)})^{n}
$$
Recall the criteria established by Equation \ref{eq0} for a morphism of primitive stock \& flow diagrams (a presheaf morphism) to represent a morphism of stock \& flow diagrams.  With $\Phi(\Link)^{*}$ representing a diagonal map $\mathbb{R} \rightarrow \mathbb{R}^{n}$, the requirement for qualifying as a morphism of stock \& flow diagrams laid out by Equation \ref{eq0} implies that we must choose flow functions $\gamma_{f_{i}},  i = 1, \cdots n$ such that
$\sum_{i=1}^{n} \gamma_{f_{i}} = \psi_{f}$

That is, the resulting flow functions $\gamma_{f_{i}}$ for each ``branch'' of the split flow need to total up to a value consistent with the original full flow $\psi_{f}$, giving $n-1$ degrees of freedom.

To split a flow into a corolla, we will replace $\Phi(f)$ in \ref{corolladecomposition} by $\mathrm{Disc_{\{ *\}}} \rightarrow (\coprod_{i=1}^{n} C^{f}_{s} , \gamma_{\bullet})$ and compute the colimit.

\end{construction}

\begin{construction}\label{con3.3}
Suppose that we want to subdivide a certain flow $f$ in a stock \& flow diagram $(X, \phi_{*})$ as in \ref{con3.2}. We will show how to do this in the case that the flow is between two stocks; the case of a half flow is similar but easier. 

Suppose that the flow appears in two corollas $\mathcal{C}_{s} = (C_{s}, \gamma_{\bullet}), \mathcal{C}_{t} = (C_{t}, \psi_{\bullet})$. By construction, we have $\psi_{f} = \gamma_{f}$. By choosing flow functions $\psi_{f_{1}} = \gamma_{f_{1}}, \cdots, \psi_{f_{n}} = \gamma_{f_{n}}$, so that 
$$
\psi_{f} = \sum \psi_{f_{i}} = \sum \gamma_{f_{i}} = \gamma_{f}
$$

we can replace the flow $f$ in the corollas with $n$ flows, using the procedure of Construction \ref{con3.2} to obtain two new corollas $\mathcal{D}_{s}, \mathcal{D}_{t}$.

We will replace $\El_{/X}$ with the category $\El'$ given by:

\begin{align*}
Mor(\El') = \{ \mathfrak{F}_{g}^{\mathcal{C}_{r}} : \mathcal{U}_{g} \rightarrow \mathcal{C}_{r}   : r \neq s \in X(\Stock), g \in X(\Flow),  g \neq f \} \\
\cup \{ \mathfrak{F}_{g}^{\mathcal{D}_{s}} : \mathcal{U}_{g} \rightarrow \mathcal{D}_{x}   :  g \in D_{x}(\Flow), x =  s, t  \} \\
\Ob(\El') = \{  \mathcal{C}_{s}, y \in (X(\Stock) - \{ s, t\}) \} \cup \{ \mathcal{U}_{g} : g \in X(\Flow)- \{f \}  \} \\
\cup  \{  \mathcal{D}_{s}, \mathcal{D}_{t} \} \cup \{ \mathcal{U}_{f} : f \in D_{s}(\Flow) \cup D_{t}(\Flow) \} 
\end{align*}

and compute the colimit to split the flows. 

\end{construction}

\begin{construction}\label{con3.4}
To end our discussion, we would like to explain the full process by which we would form a hierarchical stock \& flow diagram, starting with a stock \& flow diagram $(X, \phi_{*})$

\begin{enumerate}
\item{First, we form the category $El_{/X}$, given by Construction \ref{con1.8}}
\item{We split flows in corollas, by applying Construction \ref{con3.3}, potentially multiple times to modify the category $El_{/X}$, obtaining a new category $El'$}
\item{We apply the procedure of Construction \ref{con2.3} and Definition \ref{def2.5} to $\El'$ to construct a hierarchical stock \& flow diagram $\El' \rightarrow \StockFlow$}
\item{We compute the colimit of the aforementioned diagram}
\end{enumerate} 

For instance, for Example \ref{exam2.7}, where we are substituting for the corolla $I$ in the diagram of Example \ref{exam2.2}, we would first split the flow $Recovery$ into two flows $Recovery', RecoveryFromHNICU$, and then replace the corolla of the stock $I$ with the middle diagram in the figure, using Construction \ref{con2.3}. Finally, we would compute the colimit of the diagram from Example \ref{exam2.7}
\end{construction}

\section{Upstream-Downstream Composition of Stock \& Flow Diagrams}
\label{section: upstream downstream composition}

In this section, we consider the problem of defining an `upstream-downstream composition' of two stock \& flow diagrams as described above. To accomplish this, we will exploit the colimit decomposition of stock \& flow diagrams in terms of corollas and unit flows.

\subsection{Updating flows and links}

This subsection briefly explains how to update flows and links in a stock \& flow diagram using categorical constructions. We will first explain how to achieve this for a corolla, and then for the general case. 

\begin{construction}\label{con1.2}
Suppose that corolla $\mathcal{C}_{s} = (C_{s}, \phi_{\bullet})$ in a stock \& flow diagram $(X, \gamma) \in \Ob(\StockFlow)$. Let $L = \mathcal{C}^{f}_{s}(\Link)$ be the set of (incoming) links attached to flow $f$ in $\mathcal{C}_{s}$.  Suppose we choose a triple $(f, \psi : \mathbb{R}^{L'} \rightarrow \mathbb{R}, L' \subseteq L )$, where $f$ is a flow of $\mathcal{C}_{s} $.
We want to update the link structure in such a way that $L'$ is the set of (incoming) connected links of $f$ and the flow function is given by $\psi$. 

To do this, let $\mathcal{D}_{s}^{f} = (D_{s}^{f}, \{ \psi \})$ be a diagram as in Equation \ref{eq0}, where $D_{s}^{f}(\ct) = L' \subseteq L$. 
We then replace the summand of \ref{corolladecomposition} corresponding to $\mathcal{C}^{f}_{s}$ with the stock \& flow diagram $(\mathcal{D}_{s}^{f})$ and compute the colimit. 
 
\end{construction}

\begin{remark}\label{rmk1.3}
In order to `add' flows to the corolla, we simply add summands to the diagram of \ref{corolladecomposition} corresponding to the flows we want to add. Finally, we can also subdivide flows according to the technique described in Section \ref{section: subdividing flows}. 
\end{remark}

\subsection{Upstream-Downstream composition}

Suppose that we have stock \& flow diagrams $(X, \phi_{\bullet}), (Y, \psi_{\bullet})$. Suppose we have also chosen a 3-tuple 
$$\mathcal{T} = (a, b, f) \in X(\Stock) \times Y(\Stock) \times X(\Flow) $$
\noindent
with $f$ a half flow out of $a$ in the sense of Definition \ref{def2.1}. In this section, we will describe a stock \& flow diagram $X \Rrightarrow_{\mathcal{T}} Y$ which is obtained by `connecting' $a, b$ via the flow $f$, and then adding link structure pointing to the flow $f$. We can think of this as a variety of upstream-downstream composition. 

\begin{construction}\label{con2.1}
In this construction, we will describe the steps involved in constructing the `upstream-downstream composition' $(X, \phi_{\bullet}) \Rrightarrow_{\mathcal{T}}  (Y, \psi_{\bullet})$. 

Let $\mathcal{U}_{f} = (U_{f}, u_{\bullet})$ be the unit flow associated with $f$ in $(X, \phi_{\bullet})$. There is a unique stock \& flow diagram $\mathcal{D}_{b}^{f} = (D_{b}^{f}, u_{\bullet})$ with 

$$
D_{b}^{f}(p) =
\left\{
	\begin{array}{ll}
		*  & \mbox{if } p = \Stock, \Inflow \\
		U_{f}(p) & \mbox{if } \text{otherwise} 
	\end{array}
\right.
$$

Consider the unique diagram
$$
\xymatrix
{
(X, \phi_{\bullet}) \ar[r] & \mathcal{B} \ar[r] & \mathcal{C} \\
\mathcal{U}_{f} \ar[u]_{i_{1}} \ar[r]_{i_{2}} &  \mathcal{D}_{b}^{f} \ar[u] \ar[r] & \ar[u] \mathcal{A} \\
& \mathrm{Disc}_{\{ *\}} \ar[u]_{i_{3}} \ar[r]_{b} & \ar[u] (Y, \psi_{\bullet})
}
$$
where $i_{1}, i_{2}, i_{3}$ are the obvious inclusion maps, and $b$ takes the unique stock of $\mathrm{Disc}_{\{ *\}}$ to $b$, and all squares are pushouts. We will then define: 
$
(X, \phi_{\bullet}) \Rrightarrow_{\mathcal{T}} (Y, \psi_{\bullet}) := \mathcal{C}
$
\end{construction}

\section{Conclusion}
\label{section: conclusion}

This work has described a novel approach to model composition addressing two common problems:  Firstly, we contribute a solution to the need to expand the characterized dynamics for one particular section of a stock \& flow diagram (until now represented by a single stock) into greater detail in the form of a whole diagram.  Given the frequency with which this situation is encountered, such hierarchical composition provides a much-needed complement the previously contributed approach to lateral composition of stock \& flow models \cite{AlgebraicStockFlow, CompositionalStockFlowMfPHBook2022}.  Secondly, we provide a solution to address the frequent modeling circumstance in which one model characterizes circumstances antecedent to another, and where there are material flows from the antecedent (or upstream) model into the downstream model.  As modeling tools are expanded to exploit the modularity afforded by categorical approaches to stock \& flow diagrams \cite{ACTStockFlow2022, CompositionalStockFlowMfPHBook2022}, and are supported by a growing number of tools aimed at end-user modelers \cite{CompositionalStockFlowMfPHBook2022}, we expected the opportunity and needs to employ each of these types of composition with models will grow rapidly.

The current contribution provides a solution to these two modeling needs via colimit decomposition of a stock \& flow diagram into corollas (single stocks) and unit flows.  Such decomposition allows for the substituted diagram to be glued into compatible flows incident to the stock being expanded.  In the process of pursuing this process, this work contributes insights that may be important for achieving other forms of composition.

Working with other members of the team, the authors anticipate the addition of the hierarchical decomposition described here within the AlgebraicStockFlow framework \cite{AlgebraicStockFlow} within AlgebraicJulia \cite{AlgebraicJulia}.  We further plan to add visual user interface support for such hierarchical composition into the ModelCollab real time collaborative stock \& flow modeling based on AlgebraicStockFlow, thereby making it accessible to a vast range of modelers without the requirement of understanding the underlying category theory.

\section*{Acknowledgements}
We gratefully acknowledge the insights and feedback from Evan Patterson of the Topos Institute and John Baez of UC Riverside.  Co-author Osgood wishes to express his appreciation of support via NSERC via the Discovery Grants program (RGPIN 2017-04647) and from XZO \& SYK.

\bibliography{APLCat.bib}

\section{Appendix}

\subsection{Proof of Theorem 2.4}
\begin{proof}

It is easy to verify that there is a commutative diagram of stock \& flow diagrams
$$
\xymatrix
{
(X, \phi_{\bullet}) \ar[r]_{s_{1}} \ar[d]_{s_{2}} & (Z, \kappa_{\bullet}) \ar[d]_{p_{1}}   \\
(Y, \psi_{\bullet}) \ar[r]_{p_{2}}  & (Z \coprod_{X} Y, \gamma_{\bullet}) 
}
$$
where the maps $p_{1}, p_{2}$ are induced from the maps $p_{1}, p_{2}$ in the preceding diagram.

Consider now
$$
\xymatrix
{
(X, \phi_{\bullet}) \ar[r]_{s_{1}} \ar[d]_{s_{2}} & (Z, \kappa_{\bullet}) \ar[d]_{p_{1}} \ar[ddr] &  \\
(Y, \psi_{\bullet}) \ar[r]_{p_{2}} \ar[rrd] & (Z \coprod_{X} Y, \gamma_{\bullet}) & \\
& &  (W, \omega_{\bullet})
}
$$
By the universal property of pushouts, we can find a presheaf map that makes the following diagram commute 
$$
\xymatrix
{
X \ar[r]_{s_{1}} \ar[d]_{s_{2}} & Z \ar[d]_{p_{1}} \ar[ddr] &  \\
Y \ar[r]_{p_{2}} \ar[rrd] & (Y \coprod_{X} Z)\ar@{.>}[dr]_{e} & \\
& &  W
}
$$
We check that $e$ induces a map of stock \& flow diagrams. For $f \in W(\Flow)$, we have:
\begin{multline*}
 \sum_{x \in e(\Flow)^{-1}(f)} \gamma_{x} \circ e(\Link )^{*} \\ 
=\sum_{x \in e(\Flow)^{-1}(f)} \Big ( \sum_{y \in p_{2}(\Flow)^{-1}(x)} \psi_{y} \circ p_{2}(\Link)^{*} \Big ) \circ e(\Link)^{*} \\ 
=\sum_{x \in e(\Flow)^{-1}(f)} \sum_{y \in p_{2}(\Flow)^{-1}(x)} \Big ( \psi_{y} \circ \Big ( p_{2}(\Link)^{*} \circ e(\Link)^{*} \Big ) \Big ) \\
=\sum_{y \in (p_{2} \circ e)(\Flow)^{-1}(f)} \psi_{y} \circ (e \circ p_{2})(\Link)^{*} = \omega_{f} \\
\end{multline*}
\end{proof}

\subsection{proof of Theorem 2.11}

\begin{proof}
Each colimit can be reworked as a coequalizer (apply the dual of the construction from \cite[Theorem I.7.10]{BarrWells} ).
In particular, the colimit of the inclusion functor is the coequalizer:
$$
\xymatrix
{
\big ( \coprod_{f \in X(\Inflow)} \mathcal{U}_{f} \big ) \coprod \big ( \coprod_{f \in X(\Outflow)} \mathcal{U}_{f} \big) \ar@<1ex>[r]^{\phi_{1}} \ar@<-1ex>[r]_{\phi_{2}}  & \big ( \coprod_{f \in X(\Flow)} \mathcal{U}_{f} \big )  \coprod \big ( \coprod_{s \in X(\Stock)} \mathcal{C}_{s} \big ) \ar[r] & (X_{\new}, \phi')
}
$$

where $\phi_{1}$ is the morphism $$\mathfrak{F}_{f}^{\mathcal{C}_{s}} : \mathcal{U}_{f} \rightarrow \mathcal{C}_{s} \subseteq  \big ( \coprod_{f \in X(\Flow)} \mathcal{U}_{f} \big )  \coprod \big ( \coprod_{s \in X(\Stock)} \mathcal{C}_{s} \big ) 
$$ on each factor. The map $\phi_{2}$ includes each factor $\mathcal{U}_{f}, f \in X(\Inflow)$ ($\mathcal{U}_{f}, f \in X(\Outflow)$ ) in the summand $\mathcal{U}_{X(\into)(f)}$ ($\mathcal{U}_{X(\outto)(f)}$) of $\big ( \coprod_{f \in X(\Flow)} \mathcal{U}_{f} \big )$. Note that this colimit exists by \ref{thm1.4}.

The maps induced by inclusion $\mathcal{C}_{s} \rightarrow (X, \phi)$ yield a morphism $(X_{\new}, \phi') \rightarrow (X, \phi)$ by the universal property of coequalizers. We check that this is an isomorphism. It suffices to check that it is induces an isomorphism on primitive stock \& flow diagrams. 

Coequalizers in a presheaf category are computed pointwise, and   for $p \neq \Flow, \Link$, $U_{f}(p) = \emptyset$. 
The usual description of coequalizers in $\FinSet$ yields, for an object $p \neq \Link, \Flow$ and $i : p \rightarrow \Stock$ the unique arrow pointing to $\Stock$: 
$$
X_{new}(p) = \coprod_{s \in X(\Stock)} C_{s}(p) = \coprod_{s \in X(\Stock)} i^{-1}(s) = i^{-1}(X(\Stock)) = X(p)
$$
For any $p, q \neq \Flow, \Link$, and any morphism $\phi : p \rightarrow q$, it is clear from the discussion above that $X(\phi)$ is the disjoint union of the $C_{s}(\phi)$, so that $X(\phi) = X_{\new}(\phi).$

 We note that $X_{\new}(\Flow) =  X(\Inflow) \coprod X(\Outflow) $, modulo the equivalence relation which identifies $x, y$ if they appear as both an inflow and an outflow. But this is just $X(\Flow)$. Similarly, $X_{\new}(\Link)$ is the coequalizer of:
 $$
 \xymatrix
 {
\big ( \coprod_{f \in X(\Inflow)} U_{f}(\Link) \big ) \coprod \big ( \coprod_{f \in X(\Outflow)} U_{f}(\Link) \big) \ar@<1ex>[r]^<<<<{\phi_{1}} \ar@<-1ex>[r]_<<<<{\phi_{2}}  &  \big ( \coprod_{s \in X(\Stock)} C_{s}(\Link) \big ) 
}
 $$
 which is just $X(\Link)$. For the arrows $\phi$ pointing to $\Flow, \Link$, we can now identify $X(\phi)$ and $X_{\new}(\phi)$.

A remaining task is to show that the flow functions are identified with those of $X$. This follows from the description of pushouts in $\StockFlow$. 

\end{proof}

\subsection{An example of upstream-downstream composition}

\begin{figure}[H]
    \centering
     \includegraphics[width=0.8\textwidth]{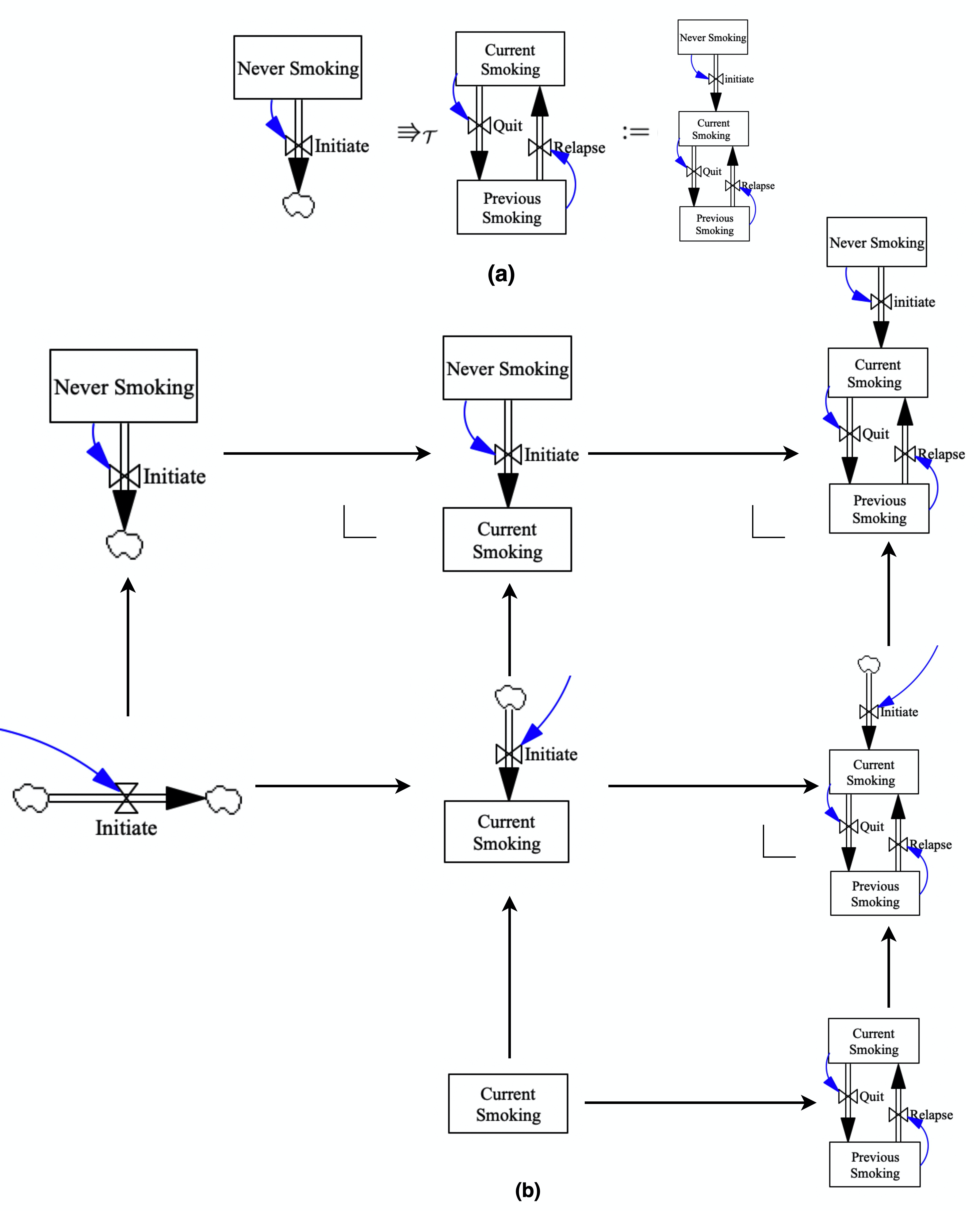}
    \caption{An example of upstream-downstream composition}
    \label{up-down-stream}
\end{figure}

\end{document}